\documentclass[12pt]{iopart}

\usepackage{amssymb}
\usepackage{graphicx}
\usepackage{relsize}
\usepackage{verbatim}
\usepackage[utf8]{inputenc} 
\usepackage{euscript}
\usepackage{amsfonts}
\usepackage{iopams}

\newcommand{\mds}[1] {\frac{d {#1}}{d\tau}}

\newcommand{\mdt}[1] {\frac{d {#1}}{dt}}

\begin{document}

\title[Proposed method of combining continuum mechanics with Einstein Field Equations]{Proposed method of combining continuum mechanics with Einstein Field Equations}

\author{Piotr Ogonowski}

\renewcommand{\baselinestretch}{1.5}

\ead{piotrogonowski@kozminski.edu.pl}
\address{Kozminski University, Jagiellonska 57/59, Warsaw, 03-301, Poland}
\vspace{10pt}

\begin{abstract}
The article proposes an amendment to the relativistic continuum mechanics which introduces the relationship between density tensors and the curvature of spacetime. The resulting formulation of a symmetric stress-energy tensor for a system with an electromagnetic field, leads to the solution of Einstein Field Equations indicating a relationship between the electromagnetic field tensor and the metric tensor. In this EFE solution, the cosmological constant is related to the invariant of the electromagnetic field tensor, and additional pulls appear, dependent on the vacuum energy contained in the system. In flat Minkowski spacetime, the vanishing four-divergence of the proposed stress-energy tensor expresses relativistic Cauchy's momentum equation, leading to the emergence of force densities which can be developed and parameterized to obtain known interactions. Transformation equations were also obtained between spacetime with fields and forces, and a curved spacetime reproducing the motion resulting from the fields under consideration, which allows for the extension of the solution with new fields.
\end{abstract}


\maketitle
\eqnobysec

\renewcommand{\baselinestretch}{1}

\section{Introduction}

Currently, field phenomena in physics is described in many ways, e.g., \cite{ft1}, \cite{ft2}, \cite{in1}, \cite{in2}, \cite{in3}, \cite{in4}, but in most field theories \cite{in5}, the field is still something additional to the spacetime - not natural consequence of spacetime existence. There are also still some challenges in describing systems that contain electromagnetic field.  Stress-energy tensor for a system with electromagnetic field \cite{EMCF}, derived from the widely accepted Lagrangian density \cite{LagDen}, is not symmetrical \cite{MEDP} and attempts are still being made to link the description of such a system with the GR, e.g. \cite{EMG1}, \cite{EMG2}, \cite{EMG3}.\\
~\\
Much theoretical work was also done to combine the equations of GR and fluid dynamics, e.g. \cite{FDG1}, \cite{FDG2}, \cite{FDG3}, \cite{FDG4}, \cite{FDG5}, however, so far the general solutions connecting these two branches of physics are unknown. There are also some unresolved problems, e.g. with the dependence of four-velocity on four-position. In relativistic electrodynamics, it is often assumed that the four-velocity is independent of the four-position, while a large number of fluid dynamics equations operate on velocity gradients, such as Navier-Stokes equations \cite{NSE} and many others.\\
~\\
The motivation of this article was to find a general solution to the Einstein Field Equations that would explain electrodynamics in curved spacetime, allow for generalization to other fields and be consistent with the equations of the continuum mechanics. The article may also be considered as the voice in still present scientific discussion about foundations of electromagnetism and its relation to spacetime geometry and spacetime itself, discussed e.g. in \cite{em1}, \cite{em2}, \cite{em3}, \cite{em4}, \cite{em5} and \cite{em6}.\\
~\\
In the first part of the article, the consequences of Hamiltonian mechanics for electrodynamics were considered. The conclusions were then used to make a minor tweak to relativistic continuum mechanics equations. Finally, the symmetric stress-energy tensor was proposed for a system containing an electromagnetic field, and then it was used to analyze the transformation to curvilinear coordinates and its relation to Einstein Field Equations. \\
~\\
The author uses the Einstein summation convention, metric signature $(+,-,-,-)$ and some standard definitions: t denotes coordinate time, $\tau$ denotes test body proper-time, m denotes test body rest mass, q denotes test body charge, S denotes Hamilton's principal function (action), L denotes Lagrangian, $\mathcal{L}$ denotes Lagrangian density, H denotes Hamiltonian.\\
~\\
The author also uses some standard four-vector definitions: $U^{\alpha}$ for four-velocities, $P^{\alpha}$ for four-momentums, $F^{\alpha}$ for four-forces, $\mathbb{F}^{ \alpha \beta} $ for electromagnetic tensors, $A^{\alpha}$ for four-accelerations, $\mathbb{A}^{\alpha} \equiv \left( \frac{ \phi }{c}, \vec{\mathbb{A}} \right )$ for electromagnetic four-potentials, $J^{\alpha}$ for four-currents, $H^{\alpha} \equiv \left ( \frac{H}{c} , \vec{p_h} \right )$ for generalized, canonical four-momentums.\\

\section{From Hamiltonian mechanics to geometry of spacetime} \label{1Chap}

One may start discussion considering Lagrangian and Hamiltonian mechanics \cite{LHMe} in flat Minkowski spacetime. Using Hamilton–Jacobi equations, one may express generalized canonical four-momentum $H^{\alpha}$ as a function of Hamilton's principal function $S$ \cite{HPF} as follows
\begin{equation} \label{EF0}
H^{\alpha} \equiv \left ( \frac{H}{c} , \vec{p_h} \right ) = -\partial^{\alpha} S
\end{equation} 
For a system containing only electromagnetic field above takes form of  
\begin{equation} \label{EF1}
H^{\alpha} = P^{\alpha} + q\mathbb{A}^{\alpha}
\end{equation} 
where  $\mathbb{A}^{\alpha}$ is the electromagnetic four-potential. This equation yields the relativistic Lagrangian \cite{HPF} (minimal coupling) for the electromagnetic field
\begin{equation} \label{EF4}
- L =  \frac{1}{\gamma} \cdot  U_{\alpha}H^{\alpha} = mc^2 \frac{1}{\gamma} + q(\phi - \vec{u}\vec{\mathbb{A}})
\end{equation}
It is known relativistic version of Lagrangian and Hamiltonian for electromagnetism, however, there is something that was missed what is oversight of important consequences. Taking four-gradient on (\ref{EF1}) for both indexes and subtracting from each other, one obtains
\begin{equation} \label{EF5}
\partial^{\beta} P^{\alpha} - \partial^{\alpha}P^{\beta} = q (\partial^{\alpha} \mathbb{A}^{\beta} - \partial^{\beta} \mathbb{A}^{\alpha})
\end{equation} 
Element related to $H^{\alpha}$ vanished, since $H^{\alpha}=-\partial^{\alpha} S$  and from calculus rules for any scalar $S$ there is
\begin{equation} \label{EF6}
\partial^{\beta} \partial^{\alpha} S - \partial^{\alpha} \partial^{\beta} S = 0
\end{equation} 
what is fundamental rule behind gauge fixing \cite{Gauge} for electromagnetic field. \\
~\\
Above reasoning and eq. (\ref{EF5}) shows that in considered system, four-momentum is dependent on four-position. It is also worth noting, that there are many concepts of continuum mechanics that depend on velocity gradients. An example would be Cauchy stress tensor, deviatoric stress tensor \cite{CMVT} or vorticity \cite{VOR}, which is a term from dynamical theory of fluids that describes velocity rotation of a fluid element, usually denoted as $\omega$ and defined as $\vec{\omega} \equiv \nabla \times \vec{u}$.\\
~\\
Velocity gradients and velocity gradient tensors are important concepts of fluid dynamics \cite{VGR1}, \cite{VGR2}, \cite{VGR3} thus velocity independent of the four-position would create significant problems for continuum mechanics. It would be also difficult to combine continuum mechanics with GR, discarding the key elements of continuum mechanics. \\
~\\
Therefore, for further discussion, the conclusion from (\ref{EF5}) and conclusions from the continuum mechanics will be adopted and it will be assumed that in considered system four-velocity depends on four-position. As it will be shown soon, such an assumption (after a minor amendment) does not cause problems for GR and electrodynamics, and in fact leads to the integration of these branches of physics. \\
~\\
Analyzing (\ref{EF5}) from the gauge theory perspective, in considered system (system containing only electromagnetic field), four-momentum $P^{\alpha}$ is just some chosen gauge for the electromagnetic four-potential. Electromagnetic filed tensor $\mathbb{F}^{ \alpha \beta}$ for such system may then be expressed equivalently as
\begin{equation} \label{EN9a}
\mathbb{F}^{ \alpha \beta} = \partial^{\alpha} \mathbb{A}^{\beta}  -   \partial^{\beta} \mathbb{A}^{\alpha} = \frac{1}{q} \left ( \partial^{\beta} P^{\alpha} - \partial^{\alpha}P^{\beta} \right )
\end{equation}
what produces the Lorentz force $F^{\alpha}$ by
\begin{equation} \label{EN9b}
F^{\alpha} = U_{\beta} \partial^{\beta} P^{\alpha} = qU_{\beta} \, \mathbb{F}^{ \alpha \beta} 
\end{equation}
since the Minkowski metric property gives
\begin{equation} \label{MPR}
U_{\beta}P^{\beta} = mc^2 \quad \rightarrow \quad U_{\beta}\partial^{\alpha}P^{\beta}=\frac{1}{2}\;  \partial^{\alpha} \left ( U_{\beta}P^{\beta} \right ) = 0 
\end{equation}
~\\
The four-current  $J^{\alpha}$ issue remains to be clarified, where
\begin{equation} \label{FCR}
{\mu_o} J^{\alpha} \equiv  \partial_{\beta} \, \mathbb{F}^{ \alpha \beta}
\end{equation} 
and where $\mu_o$ represents the permeability of free space. The continuity equation requires that $\partial_{\alpha} J^{\alpha}=0$. Denoting $\rho_o$ as rest charge density, it is clear, that the classical equation $J^{\alpha} = \rho_o U^{\alpha}$ requires vanishing four-divergence of $U^{\alpha}$. However, assuming $U^{\alpha}$ as dependent on four-position, one may also assume, that four-divergence of $U^{\alpha}$ does not vanish and note some inconsistency in the classical calculation of the density flux, which is clearly visible for volumetric mass density.\\
~\\
In the considered system, analyzed as a continuum, there is four-momentum density and four-current. One may thus consider $\varrho_o$ as volumetric mass density in some volume V for the system at rest
\begin{equation} \label{VMD}
\varrho_o \equiv \frac{m}{V}
\end{equation} 
Following the reasoning behind the calculation of the energy density in the stress-energy tensor \cite{SET}, it should be noted that both the mass $m$ and the volume $V$ are subject to Lorentz contraction effects ($m \to m\gamma $ and $V \to V\frac{1}{\gamma}$). In the four-momentum $P^{\alpha}$ mass is increased alone, thus contraction of the volume alone would change the density as follows
\begin{equation} \label{EN10}
\varrho = \varrho_o \gamma 
\end{equation}
For this reason, the total effect due to the increase of the mass (or charge) and volume contraction leads to four-momentum density of the form $\varrho \, U^{\alpha}$ and the four-current given by equation 
\begin{equation} \label{EN11}
J^{\alpha} = \rho \, U^{\alpha} = \rho_o \gamma \, U^{\alpha}
\end{equation}
Calculating the vanishing four-divergence of above, keeping in mind that $\gamma$ is a function of the four-position only (\ref{EN9b}), one obtains
\begin{equation} \label{EN12}
\partial_{\alpha} U^{\alpha} = - \mdt{\gamma} 
\end{equation}
~\\
The above amendment is easy to explain by analyzing the integral below
\begin{equation} \label{Uintegral}
P^{\alpha} = \frac{1}{\gamma} \int_V \varrho \, U^{\alpha} \; dV 
\end{equation}
There is no absolute rest. If it is assumed that V describes volume percieved by observer at rest, it means that the integrated matter is in motion ($\gamma \neq 1$) and thus $\varrho = \frac{dm\gamma}{dV}$. If it is assumed that matter is at rest and the integrating observer moves with $U^{\alpha}$ velocity in relation to immobile total mass m of this matter density, then the observer should integrate over contracted volume $V\frac{1}{\gamma}$, because in this case $\varrho = \frac{dm}{dV\frac{1}{\gamma}}$. This means that $ m \gamma $ will also be obtained.\\
~\\
The above reasoning would remain correct for any density in motion, providing a continuity equation for any density flux in flat Minkowski spacetime. Moreover, this amendment has very favorable ramifications for the merger with the GR. \\
~\\
If one would like to perceive the effects of the existence of a field in flat spacetime, as some form of spacetime curvature in curvilinear coordinates, then the four-divergence of $U^{\alpha}$ should vanish in curved spacetime, where the four-acceleration is replaced by the curvature of spacetime and geodesics. Therefore,
\begin{equation} \label{GU1}
U^{\alpha}_{\;\;\; ;\alpha} = 0 \quad \rightarrow \quad  \Gamma^{\alpha}_{\;\;\; \alpha \beta} U^{\beta} = \mdt{\gamma} \end{equation}
where $\Gamma^{\alpha}_{\;\;\; \alpha \beta}$ represents Christoffel symbols of the second kind. This leads to further conclusions. In flat Minkowski spacetime, four-divergence of the following tensor does not vanish
\begin{equation} \label{GU2}
 \partial_{\alpha} \, U^{\alpha} U^{\beta} = -\mdt{\gamma}U^{\beta} + A^{\beta} = \left ( 0 \, , \, \vec{a}\gamma^2 \right )
\end{equation}
where $\vec{a} \equiv \mdt{\vec{u}}$ is the classic acceleration. Its disappearance in curved spacetime thus leads immediately to the following conclusion
\begin{equation} \label{GU3}
U^{\alpha} U^{\beta}_{\;\;\; ;\alpha} = 0  \quad \rightarrow \quad  \Gamma^{\alpha}_{\;\; \alpha \mu}U^{\mu} U^{\beta} + \Gamma^{\beta}_{\;\; \alpha \mu}U^{\alpha}U^{\mu}  = - \left ( 0 \, , \, \vec{a}\gamma^2 \right ) 
\end{equation}
what, taking into account (\ref{GU1}), yields 
\begin{equation} \label{GU3b}
\Gamma^{\beta}_{\;\; \alpha \mu}U^{\alpha} U^{\mu}  = - A^{\beta}
\end{equation} 
This is the expected result, making that intrinsic covariant derivative of four-velocity vanishes in curved spacetime.
\begin{equation} \label{GU4}
\frac{D \, U^{\beta}}{D \, \tau} = \mds{U^{\beta}} + \Gamma^{\beta}_{\;\; \alpha \mu}U^{\alpha} U^{\mu}  = 0
\end{equation} 
~\\
According to above findings, the total force density $f^{\beta}$ acting in the system should be defined as follows
\begin{equation} \label{CM1}
f^{\beta} \equiv \varrho A^{\beta} = \varrho_o \gamma A^{\beta}  = \partial_{\alpha} \, \varrho \, U^{\alpha} U^{\beta} 
\end{equation}  
which is in line with the assumption behind the derivation of the Navier–Stokes equations, making it possible to derive their relativistic counterpart. \\
~\\
Finally, analyzing all above on the transition to curved spacetime for
\begin{equation} \label{GU5}
\partial_{\alpha} \, \varrho \, U^{\alpha} U^{\beta} =  f^{\beta}          \quad \rightarrow \quad       \varrho \, U^{\alpha} U^{\beta}_{\;\;\; ;\alpha} = 0  
\end{equation}
it is clear that this requires a relationship between the density tensors and some tensors describing the curvature of spacetime with vanishing covariant four-divergence, which opens the way to linking the continuum mechanics with GR. \\
~\\
Reasoning presented in this chapter opens the possibility of perceiving the presence of a field by some spacetime curvature and vice versa. Adding other fields to the system by adding to (\ref{EF1}) successive four-potentials $\mathbb{A}^{\alpha}_i $ and related constants $q_i$ of $i$-fields (marked with the $i$ index), would generalize the force equation (\ref{EN9b}) to the form of
\begin{equation} \label{EN9c}
F^{\alpha} = U_{\beta} \partial^{\beta} P^{\alpha} = \sum_i q_i \, U_{\beta} \, \left (  \partial^{\alpha} \mathbb{A}^{\beta}_i  -   \partial^{\beta} \mathbb{A}^{\alpha}_i \right )
\end{equation}
which opens up the possibility of expanding the reasoning with additional fields.\\
~\\
It is also possible to propose a solution where some interactions are the result of fluid dynamics, as presented in the next chapter. This will prove crucial for the explanation of the gravitational interaction described in GR, which cannot be described by an ordinary field four-potential.\\

\section{Results} \label{2Chap}

Analyzing the conclusions from the previous chapter, one may consider their application for the analysis of the stress-energy tensors. \\
~\\
One could build a stress-energy tensor for a system with an electromagnetic field, based on the density tensor $\varrho \, U^{\alpha} U^{\beta} $ in such a way, that the vanishing four-divergence of the stress-energy tensor would result from a cancellation of force densities.\\
~\\
The density of force due to electromagnetism $f^{\alpha}_{EM}$ in flat Minkowski spacetime may be calculated as
\begin{equation} \label{CM2}
f^{\alpha}_{EM} \equiv J_{\beta} \mathbb{F}^{ \alpha \beta} = \partial_{\beta} \left ( \eta^{\alpha \beta} \, \frac{1}{4 \mu_o}  \mathbb{F}^{ \gamma \mu} \, \mathbb{F}_{ \gamma \mu} - \frac{1}{\mu_o} \mathbb{F}^{\alpha}_{\;\;\gamma} \, \mathbb{F}^{\beta \gamma} \right )
\end{equation} 
where $ \eta^{\alpha \beta} \, \frac{1}{4 \mu_o}  \mathbb{F}^{ \gamma \mu} \, \mathbb{F}_{ \gamma \mu} - \frac{1}{\mu_o} \mathbb{F}^{\alpha}_{\;\;\gamma} \, \mathbb{F}^{\beta \gamma} $ is the classic \cite{EMSET} stress–energy tensor for electromagnetic field and where $\eta^{\alpha \beta}$ represents Minkowski metric tensor. \\
~\\
For universality and to facilitate the further analysis, one may introduce a more general definition of the stress-energy tensor for the electromagnetic field, in the form independent of the metric tensor $g^{\alpha \beta}$. For this purpose, at first step, one may introduce scalar $\Lambda_{\rho}$ with the dimension of energy density (subscript $\rho$), equal to the Lorentz invariant of the electromagnetic field tensor in the metric 
\begin{equation} \label{KPP}
\Lambda_{\rho} \equiv \frac{1}{4\mu_o} \mathbb{F}^{\alpha \mu} \, g_{\mu \gamma} \, \mathbb{F}^{ \beta \gamma } g_{\alpha \beta}
\end{equation}
Next, there will be introduced some tensor $h^{\alpha \beta}$ with the dimension of the metric tensor, defined in such a way to satisfy following properties
\begin{equation} \label{hDef}
\frac{h^{\alpha \beta}}{\frac{1}{4}h^{\alpha \beta}g_{\alpha \beta}} \equiv \frac{\frac{1}{\mu_o} \mathbb{F}^{\alpha \mu} \, g_{\mu \gamma} \, \mathbb{F}^{ \beta \gamma }}{\frac{1}{4}\frac{1}{\mu_o} \mathbb{F}^{\alpha \mu} \, g_{\mu \gamma} \, \mathbb{F}^{ \beta \gamma } g_{\alpha \beta}} = \frac{\frac{1}{\mu_o} \mathbb{F}^{\alpha \mu} \, g_{\mu \gamma} \, \mathbb{F}^{ \beta \gamma }}{\Lambda_{\rho}}
\end{equation}
\begin{equation} \label{hDef2}
h^{\alpha \beta} \, g_{\mu \beta} \, h_{\alpha}^{\; \; \mu} = 4
\end{equation}
what yields
\begin{equation} \label{hDefNew}
h^{\alpha \beta} = 2 \; \frac{\mathbb{F}^{\alpha \delta} \, g_{\delta \gamma} \, \mathbb{F}^{ \beta \gamma } }{ \sqrt{\mathbb{F}^{\alpha \delta} \, g_{\delta \gamma} \, \mathbb{F}^{ \beta \gamma } \, g_{\mu \beta} \, \mathbb{F}_{\alpha \eta} \, g^{\eta \xi} \, \mathbb{F}^{\mu}_{\; \; \xi }}}
\end{equation}
Thanks to above, using (\ref{KPP}) and (\ref{hDef}) one may define generalized (for any metric tensor $g^{\alpha \beta}$) stress–energy tensor for electromagnetic filed, denoted as $\Upsilon^{\alpha \beta}$ and defined as follows
\begin{equation} \label{CM3}
\Upsilon^{\alpha \beta} \equiv  \Lambda_{\rho} \left (g^{\alpha \beta} - \frac{h^{\alpha \beta}}{\frac{1}{4}h^{\mu \nu}g_{\mu \nu}}  \right )
\end{equation} 
which in the Minkowski spacetime will turn into the classic stress–energy tensor for electromagnetic field, mentioned in (\ref{CM2}).\\
~\\ 
Assuming that the only field in the system is an electromagnetic field and remaining in the Minkowski spacetime ($g^{\alpha \beta} \equiv \eta^{\alpha \beta}$), the below equation brings conservation of linear momentum and energy by electromagnetic interactions
\begin{equation} \label{CM4}
\partial_{\beta} \left ( \varrho \, U^{\alpha} U^{\beta} - \Upsilon^{\alpha \beta}\right ) = f^{\alpha} - f^{\alpha}_{EM} 
\end{equation} 
so this expression fits well as an expression to describe vanishing four-divergence of the stress-energy tensor for the whole system.\\
~\\
However, such stress-energy tensor would not ensure compliance with the Cauchy momentum equation \cite{CMPE1}, what is also known issue in EFE \cite{CMPE2}. Therefore to ensure this compliance and also to ensure compliance with General Relativity, one may introduce pressure $p$ in the system defined in the following way
\begin{equation} \label{MOD1}
p \equiv c^2\varrho + \Lambda_{\rho}
\end{equation} 
and define the stress-energy tensor $T^{\alpha \beta}$ for the whole system as
\begin{equation} \label{CM5}
T^{\alpha \beta} \equiv  \varrho \, U^{\alpha} U^{\beta}  - p \, \left (g^{\alpha \beta} - \frac{h^{\alpha \beta}}{\frac{1}{4}h^{\mu \nu}g_{\mu \nu}}  \right )
\end{equation} 
Vanishing four-divergence of tensor $T^{\alpha \beta}$ in flat Minkowski spacetime would now create two additional density of forces
\begin{equation} \label{KPPvd}
\partial_{\beta}T^{\alpha \beta} = 0 \quad \to \quad  f^{\alpha} - f^{\alpha}_{EM} - f^{\alpha}_{oth} - f^{\alpha}_{gr} = 0
\end{equation}
where
\begin{equation} \label{CM6}
f^{\alpha}_{oth} \equiv \frac{c^2\varrho}{\Lambda_{\rho}} f^{\alpha}_{EM} 
\end{equation}
\begin{equation} \label{gForce}
f^{\alpha}_{gr} \equiv  \frac{ \Upsilon^{\alpha \beta} }{\Lambda_{\rho}} \, \partial_{\beta} \, c^2\varrho 
\end{equation}
~\\
In the above picture, element $f^{\alpha}_{gr}$ seems to be related to the density of the gravitational force. It is not defined as interaction between bodies. This contraction of the electromagnetic stress–energy tensor expresses the phenomenon of bending the light path by the gradient of energy density. \\
~\\
Element $f^{\alpha}_{oth}$ seems to be related to the force density of other interactions, related to electromagnetism. It will be discussed in the last chapter. \\
~\\
The relationship of these force densities with the Einstein curvature tensor will be confirmed later in this chapter.\\
~\\
Now, vanishing four-divergence of tensor $T^{\alpha \beta}$ in flat Minkowski spacetime, expresses the four-dimensional relativistic Cauchy momentum equation (convective form). To see it, one may introduce tensor $\Pi^{\alpha \beta}$ defined as
\begin{equation} \label{CM14}
\Pi^{\alpha \beta} \equiv  - c^2\varrho \cdot \frac{h^{\alpha \beta}}{\frac{1}{4}h^{\mu \nu} \eta_{\mu \nu}}   
\end{equation}
Density of electromagnetic force may be expressed as 
\begin{equation} \label{CM15}
f^{\alpha}_{EM} = - \partial_{\beta} \left ( \Lambda_{\rho} \, \frac{h^{\alpha \beta}}{\frac{1}{4}h^{\mu \nu} \eta_{\mu \nu}} \right)   
\end{equation}
thus vanishing four-divergence of $T^{\alpha \beta}$, after easy rearrangement of elements, yields
\begin{equation} \label{CM16}
f^{\alpha} = \partial^{\alpha} p + f^{\alpha}_{EM} + \partial_{\beta} \Pi^{\alpha \beta}
\end{equation}
~\\
This equation expresses convective form of the relativistic Cauchy momentum equation, where $\Pi^{\alpha \beta}$ acts as a four-dimensional deviatoric stress tensor. According to present knowledge in the subject, deviatoric stress tensor depends only on velocity gradients \cite{DST}, and indeed, in the relativistic version thanks to (\ref{hDef}) and  (\ref{EN9a}), it may be expressed as
\begin{equation} \label{CM17}
\Pi^{\alpha \beta} = c^2\varrho \cdot \frac{\mathbb{Z}^{\alpha}_{\;\;\gamma} \, \mathbb{Z}^{ \gamma \beta}}{\frac{1}{4} \mathbb{Z}^{ \mu \nu} \, \mathbb{Z}_{ \mu \nu}}  \quad where \quad \mathbb{Z}^{ \mu \nu} \equiv \partial^{\mu} U^{\nu} - \partial^{\nu} U^{\mu}
\end{equation}
since all constants from (\ref{EN9a}) cancel out.\\
~\\
It means, that in the presented solution, electromagnetic interaction is described by the four-potential, while force densities $f^{\alpha}_{gr}$ and $f^{\alpha}_{oth}$ are consequences of fluid dynamics.\\
~\\
To show compliance of above with General Relativity, one may introduce three auxiliary tensors, defined for any considered metric $g^{\alpha \beta}$. At first it will be introduced tensor $R^{\alpha \beta}$ as
\begin{equation} \label{MOD2}
R^{\alpha \beta} \equiv 2 \varrho \, U^{\alpha} U^{\beta}  - p \, g^{\alpha \beta}
\end{equation}  
Next, one may define trace of this tensor by scalar R
\begin{equation} \label{MOD3}
R \equiv R^{\alpha \beta} \, g_{\alpha \beta} = - 2p - 2 \Lambda_{\rho} 
\end{equation}
and next, one may introduce tensor $G^{\alpha \beta}$ as
\begin{equation} \label{MOD4}
G^{\alpha \beta} \equiv R^{\alpha \beta} -  \frac{2R}{h^{\mu \nu}g_{\mu \nu}} h^{\alpha \beta}
\end{equation}
It is easy to calculate that
\begin{equation} \label{MOD5}
G^{\alpha \beta} - \Lambda_{\rho} \, g^{\alpha \beta} - \varrho c^2  \left (g^{\alpha \beta} - \frac{h^{\alpha \beta}}{\frac{1}{4}h^{\mu \nu}g_{\mu \nu}}  \right ) = 2 \, T^{\alpha \beta}
\end{equation}
In flat Minkowski spacetime ($g^{\alpha \beta} \equiv \eta^{\alpha \beta}$) one obtains
\begin{equation} \label{MOD6}
\partial_{\beta} \, G^{\alpha \beta} = f^{\alpha}_{gr} + f^{\alpha}_{oth}
\end{equation} 
thus in curved spacetime covariant four-divergence of $G^{\alpha \beta}$ should vanish and this tensor should be related to the curvature of spacetime corresponding to the forces $f^{\alpha}_{gr} + f^{\alpha}_{oth}$.\\
~\\
The problem of spacetime curvature may now be analyzed.\\
~\\
For this purpose, one may consider conditions where there are no forces in the system, which may be obtained by setting $p \left (g^{\alpha \beta} - \frac{h^{\alpha \beta}}{\frac{1}{4}h^{\mu \nu}g_{\mu \nu}}  \right ) = 0$ in (\ref{CM5}).  It may occur in three cases:
\begin{enumerate}
  \item lack of electromagnetic field
  \item $\varrho c^2 = - \Lambda_{\rho}$ 
  \item $g^{\alpha \beta} = h^{\alpha \beta}$
\end{enumerate}
First case is trivial, the second case is a special case of balance, thus one may concentrate on the third case in which the curved spacetime is described by the metric tensor $h^{\alpha \beta}$.\\
~\\ 
There may be doubts as to whether $h^{\alpha \beta}$ may be actually a metric tensor. Therefore, it should be noted that by definition (\ref{hDefNew}) $h^{\alpha \beta}$ is symmetrical and property $h^{\alpha \beta} \, h_{\mu \beta} \, h_{\alpha}^{\; \; \mu} = h^{\alpha \beta} \, h_{\alpha \beta} = 4$ is satisfied. It remains to check whether it meets the Bianchi identities.\\
~\\
To prove it, one may notice that vanishing $\Upsilon^{\alpha \beta}$ reduces equation (\ref{CM5}) in curved spacetime to the postulate of General Relativity
\begin{equation} \label{CM8}
T^{\alpha \beta} = \varrho \, U^{\alpha} U^{\beta} \quad \to \quad  {T^{\alpha \beta}}_{;\beta} = 0
\end{equation} 
where the stress-energy tensor $T^{\alpha \beta}$ determines the curvature of spacetime. Since contracted Bianchi identities are equivalent to conservation of energy and momentum (vanishing covariant four-divergence of the stress-energy tensor $T^{\alpha \beta}$) therefore $h^{\alpha \beta}$ indeed must be the metric tensor for curved spacetime in which all motion occurs along geodesics.\\
~\\
It should be noted that such a metric tensor $h^{\alpha \beta}$ should actually exist, assuming there is spacetime in which all the forces are replaced by the curvature of spacetime. Therefore, for such spacetime the metric tensor will be denoted by $h^{\alpha \beta}$ and in this spacetime instead of field and forces one obtains corresponding curvature. \\
~\\
However, taking above perspective, one may conclude, that the value of $h^{\alpha \beta}$ is fixed and should not depend on the metric $g^{\alpha \beta}$ under consideration, because selecting the next $g^{\alpha \beta}$ metrics closer and closer to $h^{\alpha \beta}$, one simply reduces the forces acting in the system in favor of increasing curvature of spacetime.\\
~\\
In this perspective, flat Minkowski spacetime with fields and forces and curved spacetime with $g^{\alpha \beta} = h^{\alpha \beta}$ are two different methods of describing the same phenomenon. Therefore one may calculate value of $h^{\alpha \beta}$ using the electromagnetic field tensor, present in the Minkowski flat spacetime
\begin{equation} \label{hDefFinal}
h^{\alpha \beta} = 2 \; \frac{\mathbb{F}^{\alpha}_{\;\;\gamma} \, \mathbb{F}^{\beta \gamma}}{ \sqrt{\mathbb{F}^{\alpha}_{\;\;\gamma} \, \mathbb{F}^{\beta \gamma}\, \mathbb{F}_{\alpha}^{\;\;\mu} \, \mathbb{F}_{\beta \mu}}}
\end{equation}
and use it to change description of the system to curved spacetime with help of the obtained value of the tensor $h^{\alpha \beta}$.\\
~\\
The question then arises about the relationship of the equation (\ref{CM8}) with the main GR equation. \\
~\\
It should be then noted, that in curved spacetime with the metric tensor $g^{\alpha \beta} \equiv h^{\alpha \beta}$ tensor $R_{\alpha \beta}$ (\ref{MOD2}) describes perfect fluid, where the difference between pressure $p$ in this fluid and energy density is equal to $2 \Lambda_{\rho}$
\begin{equation} \label{CM9}
R_{\alpha \beta} = \frac{1}{c^2}\left (\left [p - 2 \Lambda_{\rho} \right] +  p  \right ) U_{\alpha} U_{\beta} - p \, h_{\alpha \beta}
\end{equation}  
the value of tensor $G_{\alpha \beta}$ (\ref{MOD4}) becomes
\begin{equation} \label{CM11}
G_{\alpha \beta} = R_{\alpha \beta} -  \frac{1}{2}R \, h_{\alpha \beta}
\end{equation}
and (\ref{MOD5}) reduces to
\begin{equation} \label{CM12}
G_{\alpha \beta} - \Lambda_{\rho} \, h_{\alpha \beta} = 2 \, T_{\alpha \beta}
\end{equation}
Therefore, in presented solution, $R_{\alpha \beta}$ acts as Ricci tensor and $G_{\alpha \beta}$ acts as Einstein curvature tensor, both with an accuracy of $\frac{4\pi G}{c^4}$ constant, and cosmological constant $\Lambda$ is related to the invariant of electromagnetic field tensor, calculated for this metric ($g^{\alpha \beta} \equiv h^{\alpha \beta}$)
\begin{equation} \label{CM12L}
\Lambda = - \frac{4\pi G}{c^4} \Lambda_{\rho}
\end{equation}
It would mean, that in presented solution, vacuum energy that has been sought for years \cite{VAC} is related to the electromagnetic field that fills the entire space. It would also lead to the conclusion, that cosmological constant $\Lambda$ should be taken into account in the calculation of the metric, as they propose, inter alia, authors in \cite{KTD}.\\
~\\
It is worth noting, that in curved spacetime, Einstein tensor may also be interpreted as stress-energy tensor describing perfect fluid, however this time, vacuum energy density acts as pressure
\begin{equation} \label{CM13b}
G_{\alpha \beta} =  \frac{1}{c^2}\left (\left [\varrho c^2+p \right] -  \Lambda_{\rho}  \right ) \, U_{\alpha} U_{\beta} + \Lambda_{\rho}  \, h_{\alpha \beta}
\end{equation}
By analyzing above and equations (\ref{CM11}) and (\ref{CM12}) one may notice, that 
\begin{equation} \label{CM13c}
G_{\alpha \beta} = 0  \quad \to \quad p =  - \Lambda_{\rho} \quad \to \quad R = 0 \quad \to \quad R_{\alpha \beta} = 0
\end{equation}
thus this way one obtains Schwarzschild and Kerr vacuum solutions \cite{SKSol}. \\
~\\
The above conclusions on perfect fluids and origin of the metric tensor $h^{\alpha \beta}$ can also be understood in the context of Cauchy momentum equation presented in (\ref{CM16}). Adopting the metric tensor in such a way, to eliminate deviatoric stress one indeed makes mentioned fluid perfect and all forces disappear, what should be kept when introducing other, additional fields to the above solution. 

\section{Conclusions} \label{3Chap}

Summarizing, in curved spacetime ($g_{\alpha \beta} =h_{\alpha \beta}$) the main equation of the proposed solution (\ref{CM12}) expresses the Einstein Field Equations with an accuracy of $\frac{4\pi G}{c^4}$ constant and with cosmological constant $\Lambda$ dependent on invariant of electromagnetic field tensor $\mathbb{F}^{\alpha \gamma}$
\begin{equation} \label{CosmoCon}
\Lambda =  - \frac{\pi G}{c^4 \mu_o} \, \mathbb{F}^{\alpha \mu} \, h_{\mu \gamma} \, \mathbb{F}^{ \beta \gamma } h_{\alpha \beta} = - \frac{4\pi G}{c^4} \Lambda_{\rho}
\end{equation}
where $h_{\alpha \beta}$ is the metric tensor of the spacetime in which all motion occurs along geodesics and where $\Lambda_{\rho}$ describes vacuum energy density. These EFE drive to classic Schwarzschild and Kerr vacuum solutions, as shown in (\ref{CM13c}).\\
~\\
Stress-energy tensor $T^{\alpha \beta}$ for the system in a given spacetime described by some metric tensor $g^{\alpha \beta}$ is equal to
\begin{equation} \label{MetTen}
T^{\alpha \beta} =  \varrho \, U^{\alpha} U^{\beta} - \left ( c^2 \varrho +  \Lambda_{\rho}  \right ) \left ( g^{\alpha \beta} - \xi \, h^{\alpha \beta} \right )
\end{equation}
where $c^2 \varrho$ is energy density and where
\begin{equation} \label{XiDef}
\frac{1}{\xi} = \frac{1}{4}\, g_{\mu \nu} \, h^{\mu \nu}
\end{equation}
\begin{equation} \label{KPPSec}
\Lambda_{\rho} \equiv \frac{1}{4\mu_o} \mathbb{F}^{\alpha \mu} \, g_{\mu \gamma} \, \mathbb{F}^{ \beta \gamma } g_{\alpha \beta}
\end{equation}
\begin{equation} \label{hDefSec}
h^{\alpha \beta} = 2 \; \frac{\mathbb{F}^{\alpha \delta} \, g_{\delta \gamma} \, \mathbb{F}^{ \beta \gamma } }{ \sqrt{\mathbb{F}^{\alpha \delta} \, g_{\delta \gamma} \, \mathbb{F}^{ \beta \gamma } \, g_{\mu \beta} \, \mathbb{F}_{\alpha \eta} \, g^{\eta \xi} \, \mathbb{F}^{\mu}_{\; \; \xi }}}
\end{equation}
~\\
In flat Minkowski spacetime ($g^{\alpha \beta} \equiv \eta^{\alpha \beta}$) according to (\ref{CM16}) vanishing four-divergence of the proposed stress-energy tensor $(\partial_{\beta}T^{\alpha \beta} = 0)$ turns out to be relativistic Cauchy momentum equation which is the expected relationship.\\
~\\
To reproduce movement in curved spacetime, it will be more convenient to define flat Minkowski spacetime using the metric tensor given in polar coordinates
\begin{equation} \label{gMetMin}
g_{\alpha \beta} \equiv 
 \left[ \begin{array}{cccc}
1 & 0 & 0 & 0 \\
0 & -1 & 0 & 0 \\
0 & 0 & -r^2 & 0 \\
0 & 0 & 0 & -r^2sin^2(\theta) 
\end{array} \right]
\end{equation}
Total force density $f^{\alpha}$ acting in the system calculated from $\partial_{\beta} \; T^{\alpha \beta} =0$ is equal to
\begin{equation} \label{forceDen}
f^{\alpha} = \cases{
f^{\alpha}_{EM} \equiv -\Lambda_{\rho} \, \partial_{\beta} \xi \, h^{\alpha \beta} \quad (electromagnetic) \\
+ \\
f^{\alpha}_{gr} \equiv c^2 \left ( g^{\alpha \beta} - \xi \, h^{\alpha \beta}  \right ) \partial_{\beta} \varrho \quad (gravitational) \\
+ \\
f^{\alpha}_{oth} \equiv -\varrho c^2 \, \partial_{\beta} \xi \, h^{\alpha \beta} \quad (other)}
\end{equation}
~\\
One may also transform force densities from continuum description (density in the considered volume) to discrete description (point-like masses, charges, etc.)
\begin{equation} \label{FFgr}
F^{\alpha} \equiv \frac{1}{\gamma} \int_V f^{\alpha} \; dV 
\end{equation}
where $V$ denotes volume and where $1/ \gamma$ in above expressions is the result of the amendment to the continuum mechanics introduced in equations (\ref{EN11}) and (\ref{CM1}) what was shown as actually expected to keep the continuum mechanics consistent with the Lorentz transformation. \\
~\\
Considering $\varrho$ as volumetric mass density one obtains mass dependent forces where thanks to (\ref{EN10}) it may be substituted
\begin{equation} \label{GammExpress}
\partial_{\beta} \;\varrho = \varrho \; \partial_{\beta} \ln{(\gamma)}
\end{equation}
Since 
\begin{equation} \label{LamEB}
- \Lambda_{\rho} = \frac{1}{2\mu_o} \left ( \frac{E^2}{c^2} - B^2  \right) 
\end{equation}
the result of the integral of the above is unknown. One may therefore introduce a parameter $E_{\Lambda}$ with the dimension of energy 
\begin{equation} \label{consL}
\frac{1}{\gamma} E_{\Lambda}  \equiv  \frac{1}{\gamma} \int_V - \Lambda_{\rho} \; dV 
\end{equation}
Total force acting on the test body in the system may be now expressed as
\begin{equation} \label{forceDen}
F^{\alpha} = \cases{
F^{\alpha}_{EM} \equiv \frac{1}{\gamma} E_{\Lambda} \, \, \partial_{\beta} \xi \, h^{\alpha \beta} \quad (electromagnetic) \\
+ \\
F^{\alpha}_{gr} \equiv mc^2 \left ( g^{\alpha \beta} - \xi \, h^{\alpha \beta}  \right ) \partial_{\beta} \ln{(\gamma)}  \quad (gravitational) \\
+ \\
F^{\alpha}_{oth} \equiv -mc^2 \, \partial_{\beta} \xi \, h^{\alpha \beta} \quad (other)}
\end{equation}
and, according to previous section, it reproduces motion in curved spacetime given by metric tensor $h^{\alpha \beta}$ with presence of vacuum energy (non-zero cosmological constant).\\
~\\
The above equations may be tested with various spacetimes described by different metric tensors $h^{\alpha \beta}$ and can also be further developed by extending the proposed stress-energy tensor and additional parametrization.\\

\section{Discussion} \label{4Chap}
The presented solution creates a coherent picture in which spacetime is in fact a way of perceiving the electromagnetic field (what also explains equation (\ref{EN9a})). This solution allows for further development, introducing additional fields, different parameterization and simple transformation between Minkowski spacetime and curvilinear reference systems. It should be noted that the proposed solution does not question the correctness of the currently existing, well-established physical theories, but rather leads to their integration, opening up a new field for further research, experimental verification and tuning. \\
~\\
The resulting description of the gravitational interaction is a solution of the Einstein Field Equations, reproduces GR with cosmological constant $\Lambda$, complies with equations of continuum mechanics and adds components that may help explain phenomena that cannot be described with GR today. This description of gravity is also open for parameterization, development and further study of this approach in search of explanation of cosmological issues. Perhaps description of forces present in the article and the possibility of the dual description of the movement (curved spacetime vs. flat spacetime with fields and forces) may help to explain the phenomenon described today as Dark Energy \cite{DEN1}, \cite{DEN2} or explain why some fast-orbiting bodies in selected galaxies may feel less repulsive force \cite{GXRC1}, \cite{GXRC2} reducing, at least in part, the need to use Dark Matter \cite{DMI1}, \cite{DMI2}, \cite{DMDE} in the system description. It may also help with unexplained phenomena related to very massive objects that elude the currently used description of gravity \cite{UNEX0}, \cite{UNEX1}, \cite{UNEX2}, \cite{UNEX3} or help with explanation of Hubble tension problem \cite{HUBB1}, \cite{HUBB2}, \cite{HUBB3}. \\
~\\
The author intentionally does not perform the parameterization on his own, because his intention is not to create a theory explaining all the contemporary challenges of physics, but only to add his own brick to the whole knowledge by creating coherent framework that will allow the broad scientific community for further theoretical and experimental research.\\ 
~\\
It is also necessary to discuss the force density $f^{\alpha}_{oth}$ that occurs naturally in the equation (\ref{CM6}). This force density, interpreted here as "other interactions", seems to be related to strong interactions, or sum of strong and weak interactions, what would link both phenomena with additional electromagnetic force density moderated by the density of energy. This is supported by the observation that on small scales with high energy density, the density of this force will be extremely great - one may recognize it as a strong interaction property. On larger scales with small energy density, this force will be extremely weak - one may recognize it as a weak interaction property. It is also known that both of these interactions on quantum level are to some extent related to electromagnetism (charged quarks or bosons). Also the relation between strong forces and gravity has already been noted by the double copy theory \cite{DC1}, \cite{DC2}, \cite{DC3}, which may be seen as a consequence of the equation (\ref{MOD6}) vanishing in curved spacetime.\\
~\\
Due to the lack of equations describing the weak and strong fields in classical field theory, confirmation of the proposed relationship of these fields with force density $f^{\alpha}_{oth}$ must take place on the basis of quantum theories, where equation (\ref{CM6}) is a quantitative prediction that can be verified or expanded with additional components in the proposed stress-energy tensor. It also creates a new area of research to confirm the above approach or for further analysis of weak and strong interactions based on classical field theory by developing the proposed solution. \\
~\\
Finally, it is worth noting, that cosmological constant $\Lambda$ in above solution is certainly not “Einstein's greatest mistake”, but appears to be a measure for the value of invariant of the electromagnetic field tensor. Since electromagnetic field fills each considered volume regardless of its selection (from the scale of the atom to the entire space), it turns out to be a surprisingly natural explanation to the vacuum energy problem. It also may be further parameterized and extended with invariants of other fields introduced to the above solution. \\

\section{Statements} \label{5Chap}

~\\
Data sharing is not applicable to this article, as no datasets were generated or analyzed during the current study.\\
~\\
The author did not receive support from any organization for the submitted work.\\
~\\
The author has no relevant financial or non financial interests to disclose.\\

\section*{References}
\bibliographystyle{vancouver}
\bibliography{FieldsGR2SR}

\end{document}